\documentclass[12pt]{iopart}
\usepackage{iopams}
\usepackage{graphicx}
\begin{document}

\title[Rydberg transitions electric dipole moments]{Measurement of the electric dipole moments for transitions to rubidium Rydberg states via Autler-Townes splitting}

\author{MJ Piotrowicz$^1$, C MacCormick$^{1}$, A Kowalczyk$^1$, S Bergamini$^1$, II Beterov$^2$ and EA Yakshina$^2$}
\address{$^1$Department of Physics and Astronomy, The Open University, Walton Hall, MK6 7AA Milton Keynes UK}
\address{$^2$Institute of Semiconductor Physics, Pr.Lavrentyeva
13, 630090 Novosibirsk, Russia }
\ead{c.maccormick@open.ac.uk}
\pacs{32.70.Cs, 32.80.Ee}
\begin{abstract}
We present the direct measurements of  electric-dipole moments for $5P_{3/2}\to nD_{5/2}$ transitions with $20<n<48$ for Rubidium atoms. The measurements were performed  in an ultracold sample  via observation of the Autler-Townes splitting in a three-level ladder scheme, commonly used for 2-photon excitation of Rydberg states. To the best of our knowledge, this is the first systematic measurement of the electric dipole moments for transitions from low excited states of rubidium to Rydberg states. Due to its simplicity and versatility, this method can be easily extended to other transitions and other atomic species with little constraints. Good agreement of the experimental results with theory proves the reliability of the measurement method.

\end{abstract}

\pacs{32.70.Cs, 32.80.Ee}
\maketitle

\section {Introduction}
Rydberg atoms in the ultra-cold regime have attracted a lot of
theoretical and experimental interest in recent years. The strong, long range nature of the interactions in these systems can be tailored with  high precision, making them very attractive for applications in quantum information \cite{lukin2001,saffman2010} and the study of strongly interacting many-body systems \cite{saffman2010}.
Crystal-like structures can be created where the atoms are confined in space and arranged in arbitrary geometries,
and the interactions amongst them can be controllably switched on and off  and tuned
finely. The high level of control that can be achieved in these systems
and the low level of decoherence they display has prompted several groups to study
atom-atom entanglement and quantum gate operation  using single atoms in microscopic dipole
traps \cite{grangi, saffman}.

The basis of qubit encoding and quantum logic gate operations is  reliable preparation and manipulation of the electronic states of atoms and the tuning of their interactions. The recent demonstration of  interaction-driven Rydberg blockade between  two trapped ultra-cold atoms was based on the observation of an increase of the Rabi frequency in collective one-atom excitations \cite{gaetan2009}.
Similarly, strong interactions in mesoscopic ensembles of ultracold Rydberg atoms can be exploited to create a novel state, where a number of atoms behave collectively to share a single excitation \cite{johnson2008,reetz-lamour2008}. The proof of  the coherent nature of the excitation relies on the  observation of Rabi oscillations between low excited and Rydberg levels.
The Rabi oscillation  of a single atom (or blockaded sample of atoms) between a low and a Rydberg state  is governed by the dipole moment of the transition involved. In ref. \cite{miroshnychenko2010},   Rabi oscillation of single atoms were numerically simulated, elucidating the coherent nature of the excitation and giving useful insight on the experimental limits. More generally, a comparison of the experimental measurements with theoretical calculations of dipole matrix elements is required, both for verification of the experiments and to constrain the limit of validity and the accuracy of present models.

Since analytical expressions for radial matrix elements are known for hydrogen only \cite{bethe1977}, numerous theoretical models have been developed for the calculation of  dipole matrix elements for alkali-metal atoms. Recent measurements of the effective lifetimes of Rydberg states \cite{marcassa2009,branden2010} can be used for indirect verification of the theory \cite{beterov2009}, but they do not reveal the spectroscopic features of each particular transition. Experimental measurement of the absolute transition probability between excited states of alkali-metal atoms is a challenging task \cite{huber1986}. The transitions between ground and first excited states have been studied in detail, both experimentally and theoretically. The oscillator strengths of the principal series of rubidium were measured  \cite{caliebe1979, shabanova1984}, as well as the  oscillator strengths for transitions between low excited and Rydberg states of sodium and lithium, presented in ref. \cite{baig2007,hussain2007}. Most of the experimental methods have limited accuracy \cite{marr1968}.

More precise measurements of the dipole moments for transitions between excited states of alkali-metal atoms are therefore necessary for the verification of  theoretical models. Particularly, measurements of the dipole moments for transitions between first excited and Rydberg states, which are important in many experiments with ultracold Rydberg atoms, have never been performed.

In the present work we describe a relatively simple method, similar to one previously used in \cite{chem} for molecular transitions, which allowed us to measure for the first time the dipole moments for Rb $5P_{3/2} \to nD_{5/2}$ transitions with $20\leq n \leq 48$. This method is based on the spectroscopic observation of the Autler-Townes splitting (AT) in a three-level ladder scheme \cite{rai}, which is also commonly used in electro-magnetically induced transparency (EIT) experiments with cold Rydberg atoms \cite{mauger2007, weatherill2008}.
\section{Method}

We observe EIT and AT splitting \cite{autler} in a sample of ultracold Rb Rydberg atoms by measurement of the absorption profile of the weak (probe) laser radiation scanned across the resonance $5S_{1/2} \to 5P_{3/2}$  in the presence of the stronger (coupling) laser radiation tuned to   transition $5P_{3/2}\to nD_{5/2}$, following the scheme suggested in \cite{rai, mauger2007, weatherill2008}.

\begin{figure}
 \centering
 \includegraphics[height=5cm]{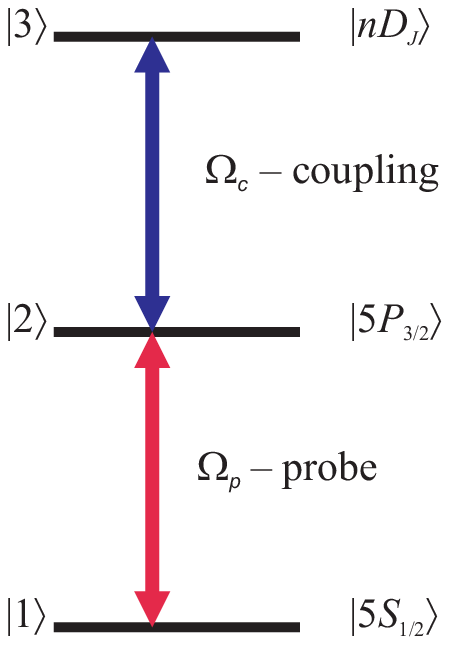}%
\caption{Schematic representation of the three-level ladder scheme. $\Omega_C$ and $\Omega_P$ are the Rabi frequencies associated to the coupling laser and probe laser respectively. \label{fig:levels_scheme}}
\end{figure}

The long-lived states $\vert 1\rangle$ and $\vert 3\rangle$ are coupled to a short lived state $\vert 2 \rangle$ by a weak probe laser and an intense coupling laser, as shown in figure \ref{fig:levels_scheme}, and we assume that the states $\vert 1\rangle$ and $\vert 3\rangle$ are not coupled by electric dipole transition.
With the atoms initially populating state $\vert 1\rangle$ and the coupling laser tuned to the $\vert 2\rangle$ to $\vert 3\rangle$ resonance, EIT appears as a dip in the center of the absorption lineshape of the probe beam, which results from the destructive interference between excitation pathways,  and is observed by sweeping the probe laser through the $\vert 1\rangle\to \vert 2\rangle$ transition. With increasing coupling laser intensity the transparency dip becomes wider, until it eventually resembles two individual lines (AT splitting), resulting from the dynamic splitting of the middle energy level due to the strong coupling field.

To obtain an expression for the absorption lineshape of the probe beam, we solve the optical Bloch equations for the system of three atomic levels and two coupling lasers, taking the probe laser as a weak perturbation so that the population of the the two states $\vert 2 \rangle$ and $\vert 3 \rangle$ are set to 0. This is ensured by keeping the probe laser intensity $I_{p}\ll I_{sat}$ for the transition to $\vert 5P_{3/2}\rangle$. We find that the cross-section for absorption of the probe laser, $\sigma_{P}$ is then:
\begin{equation}
\sigma_{P}=\sigma_{0}
\left[
\left( \left( \Gamma+2i\delta  \right)+
\frac{\Omega_{C}^2}
{\gamma_{3}+2i \left( \delta+\Delta \right)} \right)^{-1}+c.c \right]
\end{equation}
where $\sigma_0$ is the cross-section for absorption in the absence of the coupling laser, $\Gamma$ is the width of the $\vert 2\rangle$ state, $\delta$ the detuning of the probe beam, $\Delta$ the detuning of the coupling beam and $\gamma_{3}$ is the dephasing rate of the $\vert 3 \rangle$ state.  In our experiment, state $\vert 1\rangle$ corresponds to the $5S_{1/2} F=2, m_F=2$ state, $\vert 2\rangle$ to $5P_{3/2} F=3, m_F=3$ and the state $\vert 3\rangle$ to one of the Rydberg states $nD_{5/2}, F=4, m_F=4$, where n lies between 20 and 48. It is to be mentioned that other states (e.g. $nD_{3/2}$) can be also selected, by tuning the frequency and changing the polarization of coupling and probe beam.  For the $5S_{1/2} F=2\to 5P_{3/2} F=3$ transition, $\sigma_{0}=2.90 \times 10^{-13}$ m$^{-2}$.

By scanning the frequency of the probe beam around the $\vert 1\rangle \to \vert 2\rangle$ resonance, we obtain the spectrum $\sigma_{p}(\delta)$, from which we determine $\Omega_{C}$. By investigating
the dependence of $\Omega_{C}$ on the
power of the coupling beam for a variety of Rydberg states, we experimentally determine their transition dipole moments.
We find that  this  is a highly reliable and precise method of determining  dipole moments. As it will be shown in the next section, the method allows for a strongly focused coupling beam to be used, such that $\Omega_{C} \to \Omega_{C}\left(x, y\right)$, provided an accurate determination of the laser beam profile and power is performed.

\section{Experimental details}

We prepare an ensemble of cold atoms at a density of $6 \times 10^9$ cm$^{-3}$ and a temperature of 10 $\mu $K, by collecting $3 \times 10^6$ atoms from the background vapour in a magneto optical trap and allowing further cooling in optical molasses. The sample is prepared in the $\vert  5S_{1/2}, F=2, m_F=2\rangle$ magnetic sub-level by optical pumping, with an efficiency of $ 80 \%$.

\begin{figure}
 \includegraphics[width=16cm]{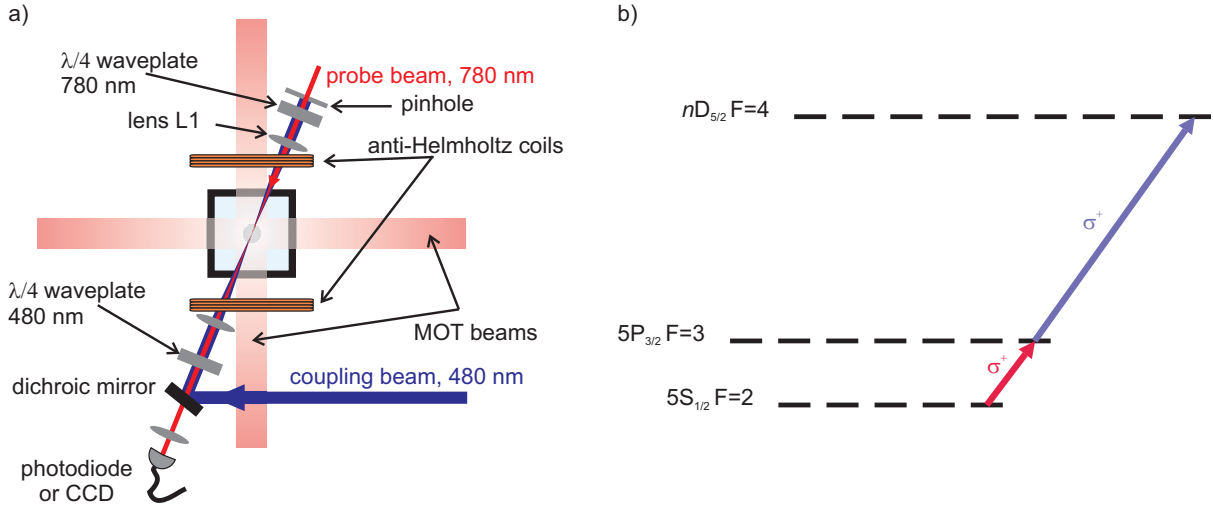}
\caption{a) Scheme of the experimental apparatus showing the MOT vacuum chamber  and the probe and coupling beams arrangement. b) Diagram of the tuning of the probe (red) and coupling laser (blue), showing the $\vert 5 S_{1/2}$, $\vert 5 P_{3/2}$, $\vert n D_{5/2}$ states and the hyperfine structure (including magnetic sublevels) \label{realscheme}}
\end{figure}

The atoms are subsequently illuminated by an intense coupling laser and a weak, counterpropagating probe laser, as shown in figure \ref{realscheme}a), which are $\sigma^{+}/\sigma^{+}$ polarized. The coupling laser is provided by a frequency doubled diode laser (TOPTICA SHG) which is frequency locked to the center of the $5P_{3/2} \to nD_{J}$ transition at around 480 nm using the scheme based on EIT described in\cite{abel2009}. The beam is focused onto the atom cloud to a  small waist in order to maximize its signature on the atomic sample.  The weak `probe' beam is provided by a commercial diode laser frequency stabilized,  which is imaged onto the cold atom cloud by a singlet lens (L1) to a waist of $364 \pm 10 \mu$m. Its frequency can be swept around the $5S_{1/2} \to 5P_{3/2}$ transition at 780.24 nm.  After passing through the atoms, a system of two lenses formed an image of the probe beam on either a CCD camera or a photodiode, as shown in figure \ref{realscheme}a).
We are able to address $nD_{3/2}$ and $nD_{5/2}$ states by tuning the laser appropriately, since the $< 1$MHz linewidth of our lasers is much smaller than the fine-structure splitting for the range of n in this study.  However, we cannot resolve the hyperfine structure of these states.  Thus, in the general case, we do not have a true three level system.  By choosing both laser beams to have $\sigma^{+}$ polarization, we drive transitions such that $\Delta m_{F}$=1, according to the scheme in figure \ref{realscheme}b). As the atoms initially populate the $\vert  5S_{1/2}, F=2, m_F=2\rangle$, the choice of polarization of the probe selects  the  $\vert 5P_{3/2}, F=3, m_{F}=3\rangle$ state. For the $nD_{5/2}$ state this scheme selects only the F=4, m$_{F}$=4 sublevel.

The atoms are illuminated for 1 ms by switching on both the 480 nm coupling and 780 nm probe laser. During this time the probe laser frequency is swept across the $5S_{1/2} F=2$ to $5P_{3/2} F=3$ transition, whilst a $200$ mG magnetic field is kept on in order to preserve the projection of the atoms' magnetic moment. Finally, we record the total laser power for each realization of the experiment on  a calibrated photodiode, accurate to better than $5\%$, also accounting for the response of the photodiode due to wavelength variation. We correct for the variations of the probe laser intensity while the frequency is swept. In this way we determine the probe absorption cross section as a function of its frequency  using Beer's law.

 It must be pointed out that the Rabi frequency $\Omega_{C}$ that appears in equation 1, is in general  a function of position $\Omega_{C} \to \Omega_{C}\left(x, y\right)$.  In many experiments, a homogeneous Rabi frequency across the sample is necessary and this requires proper shaping of the coupling laser beam. The analysis method that we employ allows us to avoid the reshaping of the beam, thus enabling us to maximize the intensities involved by focusing the laser beams.

\begin{figure}
\includegraphics[width=13cm]{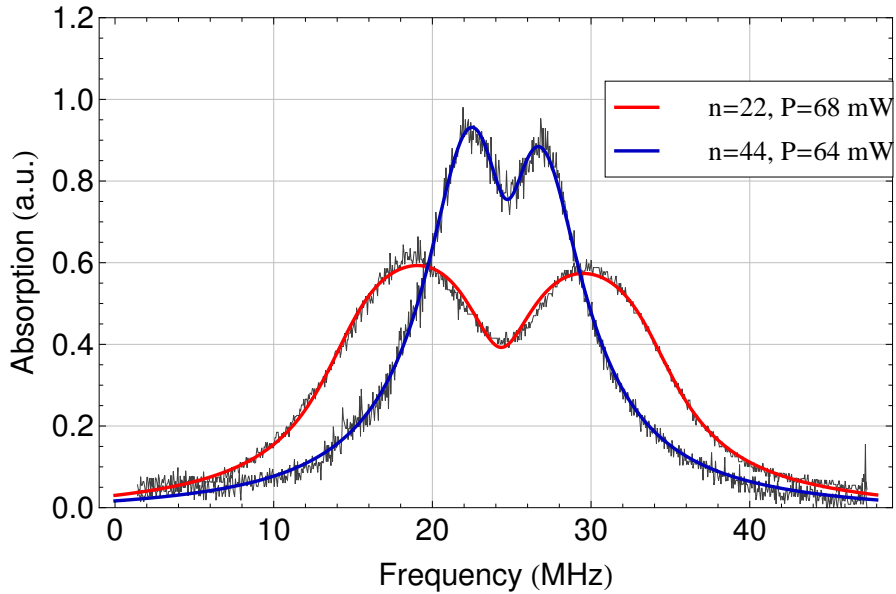}
\caption{Typical AT spectra recorded for the upper state $44D_{5/2}$ and $22D_{5/2}$. Solid lines show the modelling of the lineshape.\label{spectra}}
\end{figure}

Our method, therefore,  does not allow us to resolve the distribution $\Omega_{C}(x, y)$ of the coupling Rabi frequency due to the Gaussian intensity profile. Instead it provides an integrated signal and an `average' cross section $\sigma_{p}(\delta)$ across the spatial profile. Taking this into account in the analysis of the spectra is non-trivial as an analytic solution to equation 1 cannot be found. Thus the signal
is numerically evaluated by considering the contributions from areas of equal intensity and integrating over the beam profile, see below. By taking into account the spatial profile  $E(x, y)$ and therefore  $\Omega_{C}(x, y)$, we extract from each scan a value for $\Omega_{C}^{max}$ which correspond to the peak value of $E(x, y)$, i.e. the amplitude of the Gaussian distribution.  Figure \ref{spectra} shows how the modelling of the lineshape well adapts to the measured spectra for n=22 and n=44.

For each chosen Rydberg state $nD_{5/2}$, we take scans at different coupling powers,
while the probe laser is kept at very low intensity (typically $0.01 I_{sat}$) to minimize the probability of populating the $5 P_{3/2}$ state, as mentioned earlier.
This allows us to build a linear trend for each $n$ of $\Omega_{C}^{max}$ versus $\sqrt{P}$. From the measurements we can therefore  extract a value for the transition dipole moment for each n,  as $\Omega_{C}\propto \mu \frac{}{}\sqrt{P}$


\section{Analysis}

\begin{figure}
 \includegraphics[width=10cm]{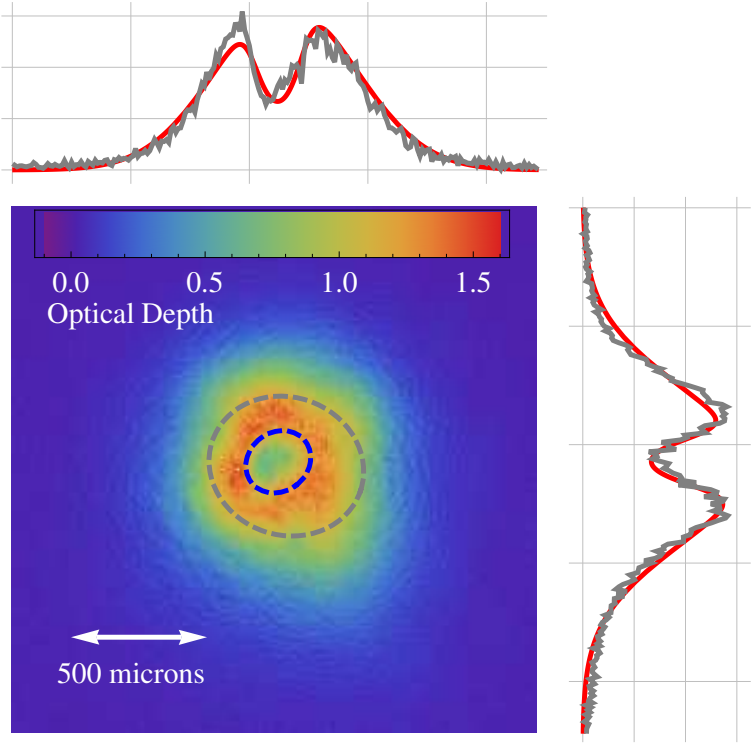}
\caption{Snap shot of the rubidium atoms absorption imaged using the probe and the counterpropagating coupling laser.  The hole in the center of the cloud is due to the reduction in cross section for the probe laser where sufficient coupling intensity exists to produce EIT.  An accurate measurement of the probe laser waist is obtained by fitting the expected 2D profile to the observed image.\label{absorption image}}
\end{figure}

We have analyzed the probe absorption lineshapes to determine the Rabi frequency $\Omega_{C}^{max}$ of the upper transition.
 For an accurate analysis, it is fundamental for us to account for the  inhomogeneity of the intensity profile of the coupling laser beam.
A useful feature of our setup is that we may take absorption images of the atoms in the presence of the coupling beam, which allows for an accurate determination of the blue laser waist at the position of the atoms.
The atomic cloud casts a shadow on the probe beam that is imaged on the camera. The presence of the coupling beam affects the optical depth of the atomic cloud, so that the cloud becomes transparent to the probe beam. By imaging the center of the atomic cloud, the optical depth of the atoms in the region illuminated by the coupling beam should reflect its intensity profile. Figure \ref{absorption image} shows a typical variation in the intensity of the light transmitted through the sample.
This allowed us to accurately determine the elliptical coupling laser waists as $w_{maj}=(240 \pm 10) \mu$m  $w_{min}=(172  \pm 10) \mu$m, along the major and minor axes of the beam cross section. These measurements are mainly limited by the resolution and signal to noise ratio of our imaging system. To make sure that the measurement were not biased, we repeated the measurements for different coupling laser powers ($80$ mW and $50$ mW) and tuning to different n states (n=26, 40, 44). We derived consistent waists in all cases, thus confirming the effectiveness of the method.
Using this determination of the laser beam profile, and the measured total power of the beam, we deduced the electric field amplitude $\bf{E}$$(x, y)$ of the coupling laser as a 2-D Gaussian distribution of amplitude $E_{max}$ and waists $w_{maj}$ and $w_{min}$, which is then taken into account via numerical methods in the analysis of the spectra.


 In order to obtain good AT spectra, we were obliged to work with very low probe intensity, and consequently we compromised the bandwidth of the photodiode amplifier to $35$ kHz in exchange for a higher signal to noise ratio. As a result, our spectra are instrumentally broadened and we are unable to observe the narrow EIT  features as seen in \cite{weatherill2008}. We quantified the broadening by analysing the spectrum obtained when the coupling laser is absent – i.e. by observing the two level $\vert 5S_{1/2} F=2\rangle \to \vert 5P_{3/2}, F=3\rangle$ transition. We find that the linewidth $\Gamma$ is $9$ MHz, larger than the value of $6.065$ MHz obtained in high precision measurements. To explain the observed broadening, we modelled the experimental signal by convolving the ideal spectrum with the response of a passive low pass filter with a corner frequency of $35$ KHz. We found that the photodiode bandwidth, combined with the measured laser linewidth of $450$ kHz, was sufficient to explain the signal broadening. This instrumental limitation determines the width of the EIT feature that we can observe, and limits our measurements of $\gamma_{3}$ to $2.5$ MHz, much larger than the values expected from the lifetime of any the Rydberg states used. In order to check that the instrumental broadening did not alter the value of $\Omega_C$ obtained from fitting the lineshape, we studied the effect of the filter on the AT signal (taking into account the inhomogeneity of the coupling laser). We found that the our fitting procedure yielded the correct $\Omega_{C}^{max}$ to high accuracy (better than $1\%$).

We analysed the spectra as follows. In general, the photodiode signal is the given by Beer's law as the integral over the atomic cloud:
\begin{equation}
S=I_{0}\int dxdydz \exp(-\sigma_{p}(x,y)n(x,y,z))
\end{equation}
where the cross-section $\sigma_{p}(x,y)$ (equation 1) is a function of x and y due to the inhomogeneity of the coupling laser and the atomic density $n(x,y,z)$. We approximate this integral as a sum using as input the measured waist sizes of the coupling beam and the measured cloud dimensions. The spectra are fitted to this sum and we obtain the parameters $\Omega_{C}^{max}$ and $\gamma_{3}$ and $\Delta$, having already determined $\Gamma$ as above. Figure 4 shows examples of the data and the fitted lineshapes.

 \begin{figure}

 \includegraphics[width=13cm]{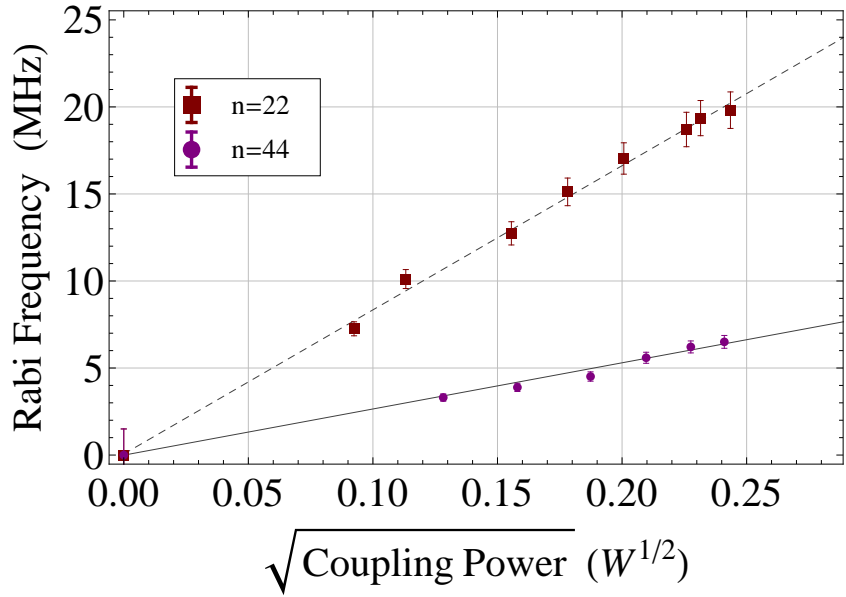}
\caption{The dependence of the Rabi frequency $\Omega$ on the square root of the coupling power $\sqrt{P}$ is used to check the inhomogeneity of the coupling laser has been correctly accounted for.  Error bars represent the uncertainty in the value of $\Omega$ obtained from the fit. \label{linearvariation}}
\end{figure}

We repeated the measurements for each n state at different coupling laser powers, ranging from $5$ mW to $80$ mW.
The results obtained for  $n=22$ and $n=44$ are plotted in figure \ref{linearvariation} together with their uncertainty, which is given from the statistics of the fitting procedure, and show a good linear trend.
The Rabi frequency is in fact directly linked to the square root of the laser power, according to
$\Omega_{n}= \mu \frac{E}{\hbar} = \mu_{n}\frac{2}{\hbar\sqrt{\pi w^2 c\epsilon_{0}}} \sqrt{P}$, where $E$ is the amplitude of the electric field in a Gaussian beam of power P and waist $w$. In our case, because of the ellipticity of the beam, $w^2 = w_{maj}\times w_{min}$.

The dipole moments for each transition  $\vert 5 P_{3/2}, F=3, m_{F}=3\rangle \to \vert n D_{5/2}, F=4, m_{F}=4\rangle$ (with n=22 to 44) can be therefore obtained directly from the gradients of the linear trends in graph \ref{linearvariation}, provided an accurate measurement of $w$ and $P$ is made.  From the linear fits we obtained the values for the gradients with an uncertainty of less than $2\%$ in most cases, with a maximum of $4\%$ for n=44. However, our uncertainties on the measurements of the waist and of the laser power were also to be taken into account when determining the values of the dipole moment $\mu_{n}$ for the transitions involved, and the total uncertainty on the measured dipole moments results to be of less than $10\%$ for all n.
\begin{figure}

 \includegraphics[width=13cm]{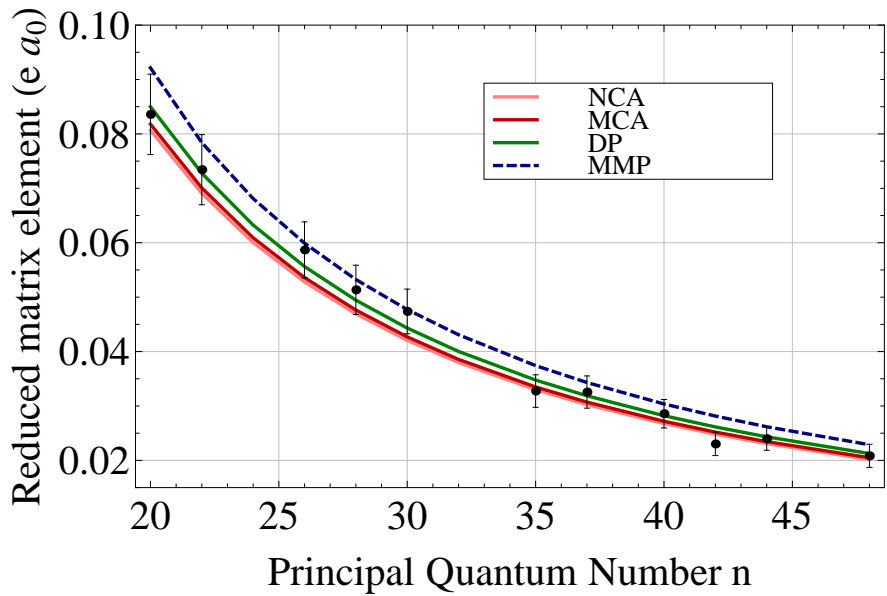}
\caption{Our results, summarized as a graph of measured reduced matrix elements $\mu$ versus principle quantum number n.  Good agreement with 3 of the 4 models is obtained, however the method of Marinescu yields slightly higher values for $\mu$ \label{results}}
\end{figure}

In figure \ref{results}  our results are plotted, after being rescaled, as reduced matrix elements to be representative of the $\vert 5P_{3/2}\rangle \to \vert nD_{5/2}\rangle$ transition, by taking into account the three-J and six-J symbols to obtain an angular coefficient of $\sqrt{\frac{2}{3}}$. The plot shows the reduced matrix elements and their uncertainties versus $n$-number.

To demonstrate the validity of our measurement method, we compared our results  to theoretical models.
Many of the methods currently used for calculations of transition dipole moments include a relatively simple numeric integration of Schr\"{o}dinger equation in the Coulomb approximation. Quantum defects are taken as input parameters  or  model potentials \cite{marinescu1994} are introduced to account for the interaction of the Rydberg electron with the atomic core.
One of the most popular methods uses the Coulomb approximation with the quantum defects of the Rydberg states taken as input parameters. The Shr\"odinger equation can be solved numerically \cite{zimmerman1997} (Numeric Coulomb Approximation, NCA). The radial matrix elements can also be obtained analytically from the quasiclassical approximation \cite{kaulakys1995,dyachkov1994} (Kaulakys or Dyachkov-Pankratov DP model) or by extrapolation of the exact radial integral for hydrogen to non-integer quantum numbers \cite{klarsfeld1989} (Modified Coulomb Approximation, MCA). A second group of methods is based on the use of the realistic model potentials (Method of model potential, MMP), which takes into account the penetration of the Rydberg electron into the atomic core, the polarization and perturbation of the core.  More complicated relativistic methods, based on Dirac equation with corrections to core polarization were used for calculation of oscillator strengths for alkali-metal principal series and transitions between excited states with  $n \leq 11$ \cite{migdalek1998,safronova2004}.  These are the most accurate and the anomalies in the ratios of experimentally measured oscillator strengths in potassium, rubidium and cesium were correctly described \cite{migdalek1998}.
Comparison of the calculated reduced dipole matrix elements \cite{loudon2000} with the experiment is shown in Fig. 4.  We calculated the dipole moments using DP , NCA , MCA  and MMP  methods, described above, using the quoted values for quantum defects for Rubidium  from  \cite{li2003}, and the  energy of the $\vert 5P_{3/2}\rangle$ state from \cite{sansonetti2006} and the model potential of Marinescu \cite{marinescu1994} with corrections for fine structure and core perturbation. For MMP calculations the RADIAL program, discussed in \cite{salvat1995}, was used. We used a `chi-squared' statistical parameter to find which model describes the experiment better. The best agreement between experiment and theory is observed for MCA model, but agreement with DP, NCA is also evident. Our results highlight a discrepancy with MMP calculations, assuming that the inherent limit in accuracy of this model is better than $5\%$, as quoted in previous works benchmarking the model. New measurements of the dipole matrix elements for other series in alkali-metal atoms will be of great importance for further verification of the theory.

\section{Outlook}
We performed  direct measurements of the dipole moments for transitions from the $5P_{3/2}$ state to high Rydberg state $nD_{J}$ for $^{87}$Rb using EIT/AT in a three-level ladder excitation scheme. These are, to our knowledge, the first direct measurements of the alkali-atom $5P_{3/2}\to nD_{5/2}$ Rydberg state dipole moments using this method. Good agreement between experiment and theory is observed,
and the measurements could be extended to different atomic levels and species, so that this method could prove invaluable for testing current theoretical models with limited accuracy.
\ack
We thank I. Ryabstev, J. Shaffer and C. W. Mansell for helpful discussions
This work was supported by EPSRC grant EP/F031130/1. S.B. and I.B. also aknowledge the RS-RFBR cost share grant 10-02-92624 that is funding this collaboration. IB and E.Ya. are also supported by the Grant of the President of Russia MK-6386.2010.2

\section*{References}
\bibliographystyle{unsrt}
\bibliography{bib}

\begin{thebibliography}{10}

\bibitem{lukin2001}
M.~D. Lukin, M.~Fleischhauer, R.~Cote, L.~M. Duan, D.~Jaksch, J.~I. Cirac, and
  P.~Zoller.
\newblock {\em Phys. Rev. Lett.}, 87:037901, 2001.

\bibitem{saffman2010}
M.~Saffman, T.~G. Walker, and K.~M\o{}lmer.
\newblock {\em Rev. Mod. Phys.}, 82:2313--2363, 2010.

\bibitem{grangi}
C~Evellin J Wolters Y Miroshnychenko P~Grangier T~Wilk, A~Gaetan and
  A~Browaeys.
\newblock {\em Phys. Rev. Lett.}, 104:010502, 2010.

\bibitem{saffman}
XL~Zhang A T Gill T Henage T A Johnson T G~Walker L~Isenhower, E~Urban and
  M~Saffman.
\newblock {\em Phys. Rev. Lett.}, 104:010503, 2010.

\bibitem{gaetan2009}
A.~Ga\"{e}tan, Y.~Miroshnychenko, T.~Wilk, A.~Chotia, M.~Viteau, D.~Comparat,
  P.~Pillet, A.~Browaeys, and P.~Grangier.
\newblock {\em Nature Phys.}, 5:115--118, 2009.

\bibitem{johnson2008}
T.~A. Johnson, E.~Urban, T.~Henage, L.~Isenhower, D.~D. Yavuz, T.~G. Walker,
  and M.~Saffman.
\newblock {\em Phys. Rev. Lett.}, 100(11):113003, 2008.

\bibitem{reetz-lamour2008}
M.~Reetz-Lamour, J.~Deiglmayr, T.~Amthor, and M~Weidem\"uller.
\newblock {\em New J. Phys.}, 10(4):045026, 2008.

\bibitem{miroshnychenko2010}
Y.~Miroshnychenko, A.~Ga\"{e}tan, C.~Evellin, P.~Grangier, D.~Comparat,
  P.~Pillet, T.~Wilk, and A.~Browaeys.
\newblock {\em .}, 2010.

\bibitem{bethe1977}
Hans~A. Bethe and Edwin~E. Salpeter.
\newblock Springer, Berlin, 1st edition, 1977.

\bibitem{marcassa2009}
Luis~Gustavo Marcassa.
\newblock {\em Physica Scripta}, 2009(T134):014011, 2009.

\bibitem{branden2010}
D.~B. Branden, T.~Juhasz, T.~Mahlokozera, C.~Vesa, R.~O. Wilson, M.~Zheng,
  A.~Kortyna, and D.~A. Tate.
\newblock {\em J. Phys. B}, 43(1):015002, 2010.

\bibitem{beterov2009}
I.~I. Beterov, I.~I. Ryabtsev, D.~B. Tretyakov, and V.~M. Entin.
\newblock {\em Phys. Rev. A}, 79(5):052504, May 2009.

\bibitem{huber1986}
M~C~E Huber and R~J Sandeman.
\newblock {\em Reports on Progress in Physics}, 49(4):397, 1986.

\bibitem{caliebe1979}
E.~Caliebe and K.~Niemax.
\newblock {\em J. of Phys. B}, 12(2):L45, 1979.

\bibitem{shabanova1984}
L.~N. Shabanova and A.~N. Khlyustalov.
\newblock {\em Opt. Spectrosc.}, 56:128, 1984.

\bibitem{baig2007}
M.~A. Baig, S.~Mahmood, M.~A. Kalyar, M.~Rafiq, N.~Amin, and S.~U. Haq.
\newblock {\em The European Physical Journal D - Atomic, Molecular, Optical and
  Plasma Physics}, 44:9--16, 2007.

\bibitem{hussain2007}
Shahid Hussain, M.~Saleem, and M.~A. Baig.
\newblock {\em Phys. Rev. A}, 75(2):022710, Feb 2007.

\bibitem{marr1968}
G.~V. Marr and D.~M. Creek.
\newblock {\em Proceedings of the Royal Society of London. Series A.
  Mathematical and Physical Sciences}, 304(1477):245--254, 1968.

\bibitem{chem}
E.~Ahmed et~al.
\newblock {\em The Journal of Chemical Physics}, 124:084308, 2006.

\bibitem{rai}
T~Cubel J R Guest P R~Berman B~K~Teo, D~Feldbaum and G~Raithel.
\newblock {\em Phys. Rev. A}, 68:053407, 2003.

\bibitem{mauger2007}
S.~Mauger, J.~Millen, and M.~P.~A. Jones.
\newblock {\em J. Phys. B}, 40(22):F319, 2007.

\bibitem{weatherill2008}
K.~J. Weatherill, J.~D. Pritchard, R.~P. Abel, M.~G. Bason, A.~K. Mohapatra,
  and C.~S. Adams.
\newblock {\em J. Phys. B}, 41(20):201002, 2008.

\bibitem{autler}
C~H~Townes S~H~Autler.
\newblock {\em Phys. Rev.}, 100:703, 1955.

\bibitem{abel2009}
R.~P. Abel, A.~K. Mohapatra, M.~G. Bason, J.~D. Pritchard, K.~J. Weatherill,
  U.~Raitzsch, and C.~S. Adams.
\newblock {\em Appl. Phys. Lett.}, 94(7):071107, 2009.

\bibitem{marinescu1994}
M.~Marinescu, H.~R. Sadeghpour, and A.~Dalgarno.
\newblock {\em Phys. Rev. A}, 49(2):982--988, Feb 1994.

\bibitem{zimmerman1997}
M.~L. Zimmerman, M.~G. Littman, M.~M. Kash, and D.~Kleppner.
\newblock {\em Phys. Rev. A}, 20(6):2251--2275, Dec 1979.

\bibitem{kaulakys1995}
B~Kaulakys.
\newblock {\em J. Phys. B}, 28(23):4963, 1995.

\bibitem{dyachkov1994}
L.~G. Dyachkov and P.~M. Pankratov.
\newblock {\em J. Phys. B}, 27(3):461, 1994.

\bibitem{klarsfeld1989}
S.~Klarsfeld.
\newblock {\em Phys. Rev. A}, 39(5):2324--2332, Mar 1989.

\bibitem{migdalek1998}
Jacek Migdalek and Yong-Ki Kim.
\newblock {\em J. Phys. B: At., Mol. Opt. Phys.}, 31(9):1947, 1998.

\bibitem{safronova2004}
M.~S. Safronova, Carl~J. Williams, and Charles~W. Clark.
\newblock {\em Phys. Rev. A}, 69(2):022509, Feb 2004.

\bibitem{loudon2000}
Rodney Loudon.
\newblock {\em The Quantum Theory of Light}.
\newblock Oxford University Press, 3rd edition, 2000.

\bibitem{li2003}
Wenhui Li, I.~Mourachko, M.~W. Noel, and T.~F. Gallagher.
\newblock {\em Phys. Rev. A}, 67(5):052502, May 2003.

\bibitem{sansonetti2006}
J.~E. Sansonetti.
\newblock {\em Journal of Physical and Chemical Reference Data},
  35(1):301--421, 2006.

\bibitem{salvat1995}
F.~Salvat, J.~M. Fern\'{a}ndez-Varea, and W.~Williamson.
\newblock {\em Computer Physics Communications}, 90(1):151 -- 168, 1995.

\end{thebibliography}

\end{document}